\documentclass[11pt]{article}
\usepackage[margin=1in]{geometry}
\usepackage{amsmath, amssymb}
\usepackage{booktabs}
\usepackage{longtable}
\usepackage{enumitem}
\usepackage{natbib}
\usepackage{hyperref}
\usepackage{listings}
\usepackage{graphicx}     
\usepackage{csvsimple}    
\usepackage{siunitx}      
\usepackage{caption}
\title{Impact by design: translating \textit{Lead times in flux} into an R handbook with code}
\author{Harrison Katz \\ Forecasting, Data Science, Airbnb \\ Harrison.Katz@airbnb.com}
\date{\today}

\begin{document}
\maketitle

\begin{abstract}
This commentary translates the central ideas in \textit{Lead times in flux} into a practice ready handbook in R. The original article measures change in the full distribution of booking lead times with a normalized L1 distance and tracks that divergence across months relative to year over year and to a fixed 2018 reference. It also provides a bound that links divergence and remaining horizon to the relative error of pickup forecasts. We implement these ideas end to end in R, using a minimal data schema and providing runnable scripts, simulated examples, and a prespecified evaluation plan. All results use synthetic data so the exposition is fully reproducible without reference to proprietary sources.
\end{abstract}

\section{What the original article established}
Let $L_t(k)$ denote the discrete distribution of lead times for arrival month $t$ on integer days $k = 0, 1, \ldots, \Delta_{\max}$. The normalized L1 divergence between two monthly distributions $L_a$ and $L_b$ is
\begin{equation}
D(L_a, L_b) = \frac{1}{2}\sum_{k=0}^{\Delta_{\max}} \left| L_a(k) - L_b(k) \right|.
\end{equation}
If $D = 0.25$ then one quarter of the probability mass has effectively shifted on the support. The article constructs $D_t$ year over year by comparing month $t$ with month $t-12$, and also relative to the same calendar month in 2018, and decomposes $D_t$ with STL to separate trend and seasonality \citep{Cleveland1990, KatzSavageColes2025}. This distributional view detects changes that means or medians may miss.

The article also provides a bound on the relative error of a pickup forecast at horizon $\Delta$:
\begin{equation}
\label{eq:bound}
|\epsilon| \le \frac{2 D \left(1 - \Delta/\Delta_{\max}\right)}{C_{\text{hist}}(\Delta)},
\end{equation}
where $C_{\text{hist}}(\Delta)$ is the historical pickup fraction at horizon $\Delta$ and $D$ is the divergence between the historical and actual lead time distributions. The bound falls as the service date approaches and grows when divergence is high or when the historical pickup fraction is small at the chosen horizon. This yields a transparent risk index that can sit next to standard pickup reporting \citep{KatzSavageColes2025}. For probability metrics background, see \citet{GibbsSu2002}.

\paragraph{Basic properties.}
The normalized $L_1$ distance equals the total variation distance on a finite support, so $D \in [0,1]$ and $D=0$ iff $L_a \equiv L_b$.
For a fixed $(D, \Delta_{\max})$, the right side of \eqref{eq:bound} is strictly decreasing in $\Delta$ and equals zero at $\Delta=\Delta_{\max}$. 
\textbf{Proof sketch.} The factor $(1-\Delta/\Delta_{\max})$ decreases linearly in $\Delta$ while $C_{\text{hist}}(\Delta)$ is nondecreasing in $\Delta$, so the quotient decreases.

\paragraph{Related modeling work for lead times}
When analysts need a parametric forecast of the entire lead time vector as a composition, the Bayesian Dirichlet auto regressive moving average model (B\textendash DARMA) models compositional time series with a Dirichlet likelihood and VARMA dynamics in additive log ratio space \citep{KatzBruschWeiss2024}. The IJF article develops B\textendash DARMA for forecasting recognition shares across future intervals and is applicable when a distributional forecast of $L_t(k)$ is desired in addition to monitoring.

\section{Why a handbook with code}
Lead time behavior is nonstationary. Most organizations already track pickup curves and a few summary metrics. Extending that reporting with two elements from the article gives leverage with minimal lift. The first element is a month by month divergence series that flags when the full lead time distribution has moved. The second element is the pickup risk bound in Equation~\eqref{eq:bound}, which converts divergence and remaining horizon into a horizon specific risk number. The handbook implements both, along with simple decision templates and an evaluation plan for use in operational settings.

\section{The R handbook}
The package \texttt{leadtimefluxR}\footnote{\url{https://github.com/harrisonekatz/leadtimefluxR}} implements signals that matter for practice and runs on local CSV files. It uses base R and common tidyverse packages and does not rely on external services.

\subsection{Minimal dataset and cohort construction}
Use one row per booking with the fields:
\begin{lstlisting}[language={}, caption=Minimal fields, label=lst:schema]
arrival_date, booking_ts, stay_nights, channel, segment, origin,
price_at_booking, cancelled, property_id
\end{lstlisting}
Arrival dates are normalized to dates. Lead time is computed in whole days and negative values are dropped. A safe default $\Delta_{\max}$ is chosen from the support and may be capped by the user. Discrete histograms $L_t(k)$ are constructed by month of arrival and by any grouping variable such as property or market. Cumulative pickup $C_{\text{hist}}(\Delta)$ is the cumulative sum of $L_t(k)$ up to $\Delta$.

\paragraph{Support and censoring.}
Pick $\Delta_{\max}$ so that at least 95\% of historical mass lies on $[0,\Delta_{\max}]$ for the group, and include a right-censored bin $\Delta_{\max}+$. 
When a nontrivial right tail exists, compute $D$ on the augmented support with a final bin $\Delta_{\max}+$ to avoid artificial spikes at the cap (cf. Figure~\ref{fig:hist}). 
For efficiency, daily bins can be aggregated beyond 28 days into weekly bins without affecting the interpretation of $D$ at horizons under 21 days.

\subsection{Divergence series and STL decomposition}
For each group and month the package computes $D_t$ either year over year or relative to 2018 by calendar month. The divergence series can be decomposed with \texttt{stats::stl} to separate trend and seasonality in a robust manner \citep{Cleveland1990}. This supports monitoring and alerting that is aligned with calendar effects.

\paragraph{Baseline choice.}
Adjacent-month $D(L_t,L_{t-1})$ is sensitive to local movements; year-over-year $D(L_t,L_{t-12})$ controls for seasonality; fixed-year baselines (e.g., 2018) stabilize long-run comparisons but can drift if the baseline regime differs structurally. 
We recommend monitoring both adjacent and YoY series, decomposed by STL, and reporting the 90th percentile of YoY $D$ as the reference level for risk mapping when at least 13 months are available.

\subsection{Pickup bound and decision templates}
The right hand side of Equation~\eqref{eq:bound} defines a horizon specific risk index. The package includes templates that map this risk to actions such as pricing cadence, advance purchase buffers, and staffing buffers \citep{TalluriVanRyzin2004}. Thresholds are explicit and can be tuned to local constraints. The approach complements model based forecasting, including B\textendash DARMA when a parametric compositional forecast is desired \citep{KatzBruschWeiss2024} and aligns with established revenue‑management forecasting practice \citep{WeatherfordKimes2003}.

\subsection{Uncertainty for $D_t$ and risk indices}\label{sec:uncertainty}
Because $D_t$ is computed from monthly histograms, finite counts can introduce variability. 
We recommend a nonparametric bootstrap at the booking level within each $(\text{group}, t)$ to obtain confidence intervals for $D_t$ and for the bound in \eqref{eq:bound}. 
For each bootstrap replicate $b=1,\ldots,B$, resample bookings with replacement within the month, rebuild $L_t^{(b)}$, recompute $D_t^{(b)}$ and the corresponding bound. 
Percentile or basic bootstrap intervals then summarize uncertainty. 
This also supports alerting rules that require $D_t$ to exceed a threshold and its lower confidence limit to exceed a smaller guardrail.

\subsection{Evaluation utilities}
The package includes MASE and sMAPE for forecast evaluation and pinball loss for quantile forecasts \citep{HyndmanKoehler2006}. Metrics are reported by horizon so that monitoring and decision rules can focus on the last two or three weeks where changes are most material.

\section{Quick start in R}
\begin{lstlisting}[language=R, caption=End to end example on synthetic data, label=lst:quick]
library(leadtimefluxR)

df <- generate_synthetic_bookings(start_date = "2021-01-01",
                                  end_date   = "2022-12-31",
                                  avg_bookings_per_day = 20,
                                  properties = 3,
                                  max_lead_days = 60,
                                  compression_level = 0.4,
                                  seed = 123)

Lk <- leadtime_histograms(df, group_cols = c("property_id"), max_lead_days = 60)
Lk_pickup <- pickup_curve(Lk, group_cols = c("property_id"))

# Divergence: adjacent always available; YoY when sample >= 13 months
D_adj <- adjacent_divergence_series(Lk, group_cols = c("property_id"))
D_yoy <- try(yoy_divergence_series(Lk, group_cols = c("property_id")), silent = TRUE)

D_est <- if (inherits(D_yoy, "try-error") || nrow(D_yoy) == 0) {
  safe_divergence_quantile(D_adj, probs = 0.90, default = 0.20)
} else {
  as.numeric(stats::quantile(D_yoy$D, probs = 0.90, na.rm = TRUE))
}

# Risk at 14 days for the latest cohort of the first property
pid <- dplyr::first(Lk_pickup$property_id)
latest <- max(Lk_pickup$month)
cohort <- subset(Lk_pickup, property_id == pid & month == latest)
Chist_14 <- cohort$Chist[cohort$k == 14][1]

bound <- relative_error_bound(D = D_est, delta = 14, delta_max = 60, Chist_delta = Chist_14)
actions <- recommend_actions(bound)
print(list(latest_month = latest, D_est = round(D_est, 3),
           Chist_14 = round(Chist_14, 3), bound = round(bound, 3), actions = actions))
\end{lstlisting}

\section{Results on synthetic data}
All artifacts are produced by the script \texttt{scripts/make\_paper\_artifacts.R} and saved under \texttt{paper\_artifacts/}.

\subsection{Adjacent month divergence}
\begin{figure}[t]
\centering
\includegraphics[width=\textwidth]{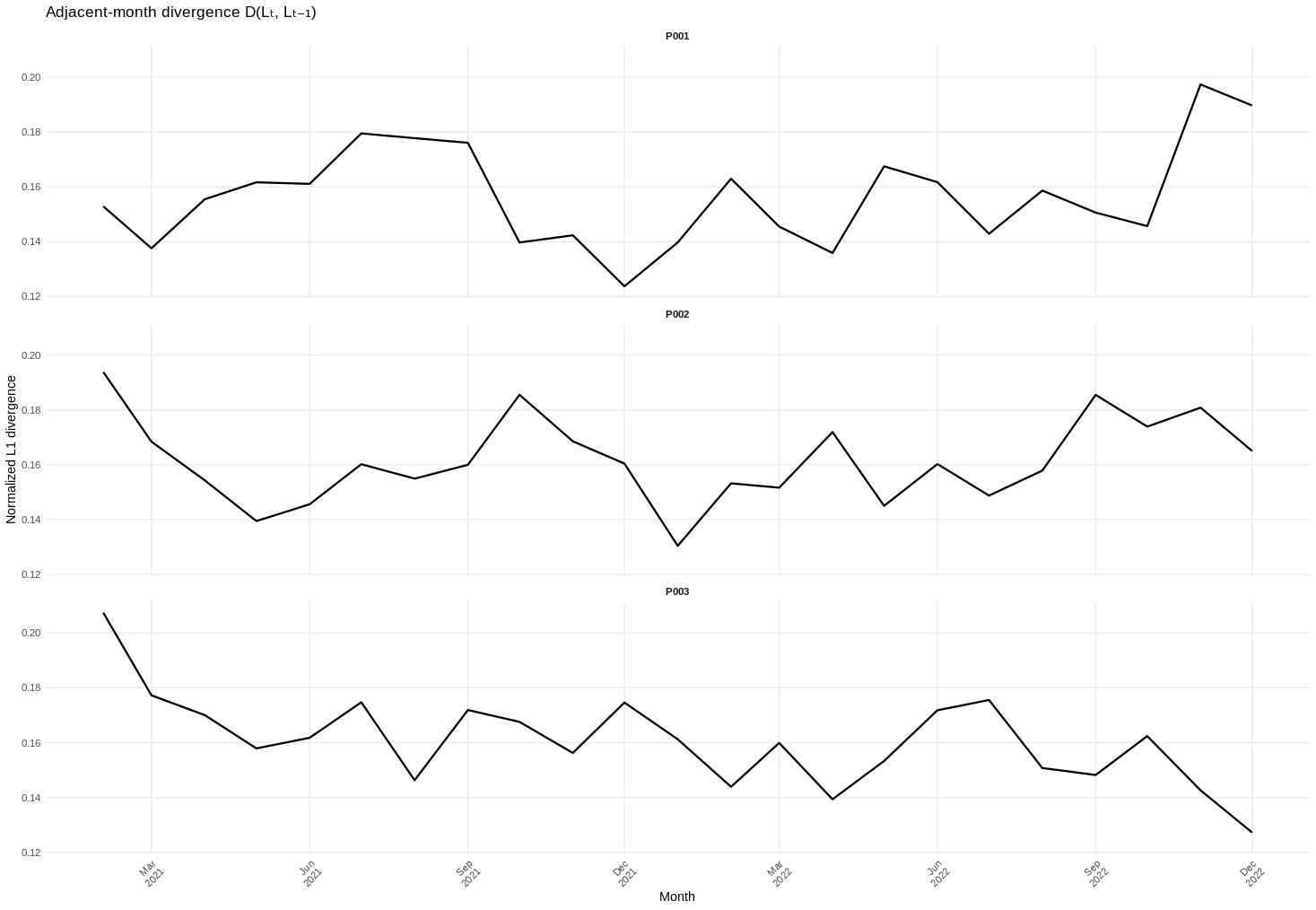}
\caption{Adjacent month divergence $D(L_t, L_{t-1})$ by property. A value of 0.20 indicates that 20\% of the mass in the lead time distribution moved relative to the prior month. Properties differ in both amplitude and pattern, which is expected under heterogeneous demand and policy environments.}
\label{fig:adj}
\end{figure}

Figure~\ref{fig:adj} shows the adjacent month normalized L1 divergence $D(L_t, L_{t-1})$ by property over 2021 to 2022 with quarterly x axis ticks for readability.

The series stay in a narrow band between roughly 0.12 and 0.20. Property P001 exhibits a late year rise, property P002 shows a dip followed by recovery, and property P003 trends downward into late 2022. These patterns are the distributional changes that a mean or a single percentile can fail to register. The 90th percentile and related summary statistics are reported in Table~\ref{tab:divsum}, pulled directly from the generated CSV.

\begin{table}[t]
\centering
\caption{Divergence summary by property (adjacent month). Values read from \texttt{paper\_artifacts/tables/tbl3\_divergence\_summary.csv}.}
\label{tab:divsum}
\begin{tabular}{lrrrr}
\toprule
Property &  Months &  Mean D &  Median D &  P90 D \\
\midrule
    P001 &      23 &   0.157 &     0.155 &  0.179 \\
    P002 &      23 &   0.162 &     0.160 &  0.185 \\
    P003 &      23 &   0.161 &     0.161 &  0.175 \\
\bottomrule
\end{tabular}
\end{table}

\subsection{Pickup and the risk bound}
\begin{figure}[t]
\centering
\includegraphics[width=0.85\textwidth]{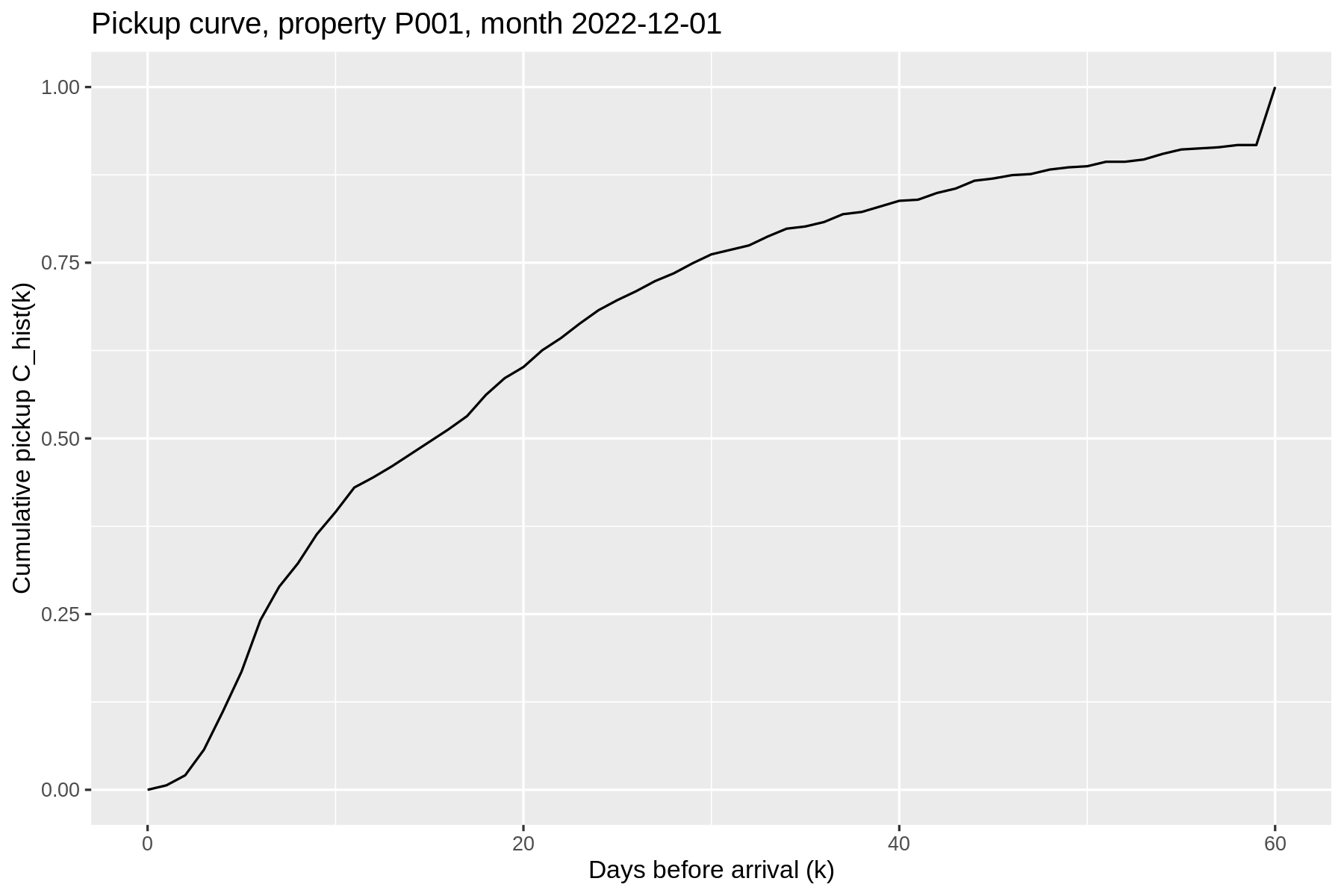}
\caption{Cumulative pickup $C_{\text{hist}}(\Delta)$ for property P001, latest month.}
\label{fig:pickup}
\end{figure}

\begin{figure}[t]
\centering
\includegraphics[width=0.85\textwidth]{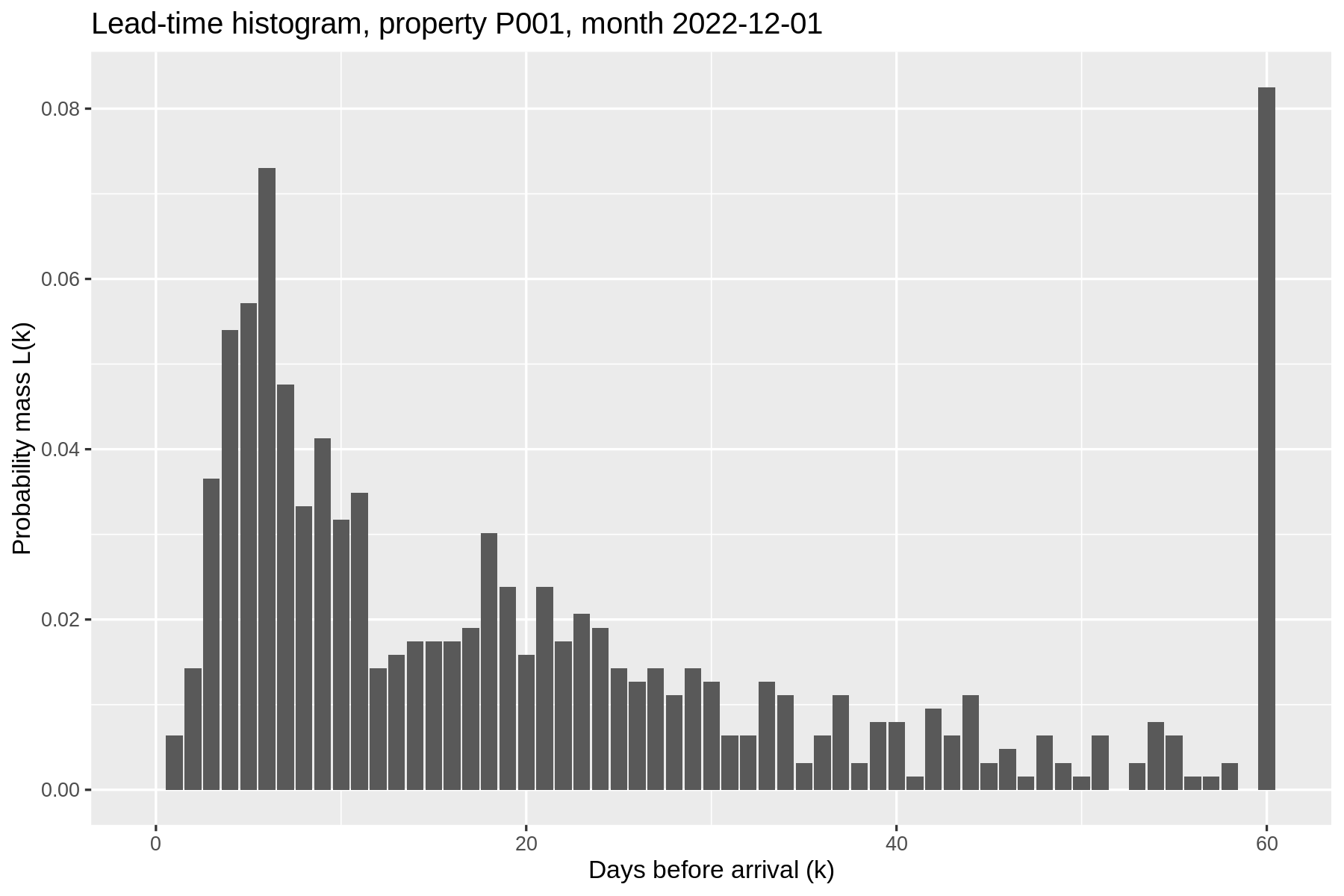}
\caption{Lead time histogram $L_t(k)$ for property P001, latest month. The tall bar at $k=60$ is a truncation artifact from capping $\Delta_{\max}$ at 60 in the simulation. In practice either increase $\Delta_{\max}$ or treat a right censored bin $60+$ separately.}
\label{fig:hist}
\end{figure}

Figure~\ref{fig:pickup} shows $C_{\text{hist}}(\Delta)$ for the latest cohort of property P001. About half of total pickup occurs by two weeks before arrival, which is consistent with the heavy short horizon mass in the corresponding histogram in Figure~\ref{fig:hist}. Combining a typical divergence in the upper teens with a 14 day horizon and $C_{\text{hist}}(14)$ near one half yields a risk bound near one half by Equation~\eqref{eq:bound}. The exact values used in the manuscript are read from the CSV in Table~\ref{tab:risk}.

\begin{table}[t]
\centering
\caption{Risk index and mapped actions at standard horizons for the latest month of each property. Values read from \texttt{paper\_artifacts/tables/tbl2\_risk\_latest\_month.csv}.}
\label{tab:risk}
\begin{tabular}{llrrrlll}
\toprule
Property &   Month &  Delta (days) &  C\_hist &  Bound & Price cadence &  AP buffer (days) & Staffing buffer \\
\midrule
    P001 & 2022-12 &             7 &   0.289 &  1.086 &      intraday &                 0 &             15\% \\
    P001 & 2022-12 &            14 &   0.478 &  0.570 &      intraday &                 0 &             15\% \\
    P001 & 2022-12 &            21 &   0.625 &  0.369 &      intraday &                 0 &             15\% \\
    P002 & 2022-12 &             7 &   0.276 &  1.138 &      intraday &                 0 &             15\% \\
    P002 & 2022-12 &            14 &   0.521 &  0.522 &      intraday &                 0 &             15\% \\
    P002 & 2022-12 &            21 &   0.634 &  0.364 &      intraday &                 0 &             15\% \\
    P003 & 2022-12 &             7 &   0.260 &  1.208 &      intraday &                 0 &             15\% \\
    P003 & 2022-12 &            14 &   0.479 &  0.568 &      intraday &                 0 &             15\% \\
    P003 & 2022-12 &            21 &   0.611 &  0.377 &      intraday &                 0 &             15\% \\
\bottomrule
\end{tabular}
\end{table}

The table reports the bound at 7, 14, and 21 days together with a simple, auditable mapping to operations. Months and properties with larger divergence and smaller $C_{\text{hist}}(\Delta)$ at the chosen horizon produce higher risk values and stricter actions. As the service date approaches the $(1 - \Delta/\Delta_{\max})$ term contracts the bound monotonically.

\section{Synthetic case studies}
To keep everything reproducible, we include calibrated simulations that mimic the empirical patterns reported in the article. A compression parameter $c \in [0, 1]$ moves probability mass toward short horizons, which raises the divergence series and the pickup risk index in predictable ways. The R scripts generate arrivals across multiple properties and segments, compute $D_t$ and $C_{\text{hist}}(\Delta)$, and apply the decision templates.

\subsection{Data generating process}
Lead times are drawn from a two component lognormal mixture. The compression parameter $c$ shifts weight toward the short horizon component and therefore increases mass inside the last two weeks. Seasonality and event weeks can be introduced as multiplicative factors so that the divergence series reflects both steady changes and occasional spikes. Segment heterogeneity is introduced as random effects. All generators accept seeds so that figures are exactly reproducible.

\subsection{Illustrative results}
As compression rises, the divergence series increases and change points appear earlier in the period. The pickup risk index is largest when divergence is high and the historical pickup fraction is low at the chosen horizon. The templates therefore suggest a higher pricing cadence and looser advance purchase buffers in those months. As the service date approaches the bound contracts by construction and the recommended actions revert to a less intensive cadence.

\section{Evaluation plan for operational settings}
Effects can be estimated with standard designs that are robust to nonstationarity.
\begin{itemize}[leftmargin=1.2em]
\item \textbf{Difference in differences}. Compare units that adopt the monitoring and decision templates against matched units on a rolling basis, using horizon specific forecast metrics as outcomes \citep{CallawaySantAnna2021}.
\item \textbf{Interrupted time series with synthetic control}. When the number of units is small, construct a counterfactual series for a treated unit and evaluate level and variance changes after the intervention \citep{Abadie2010}.
\item \textbf{Experimentation when feasible}. For pricing cadence or advance purchase rules, randomize policies at the unit level and analyze horizon specific metrics.
\end{itemize}
Primary forecast outcomes are MASE and sMAPE at 0 to 7, 8 to 14, and 15 to 21 days. Operational outcomes include spoilage, denied service, and the variance of ADR or RevPAR.

\section{Governance and reproducibility}
All analysis runs locally on booking event data. No external connections are required. The package ships with a synthetic dataset and scripts that regenerate all figures. A brief executive summary reports the divergence series, the pickup risk index at standard horizons, and any actions taken. Versioned code and seeds allow third parties to reproduce results and extend simulations.

\section{Limits}
The divergence metric captures global shifts in the distribution shape. It does not attribute causes and it does not replace demand models. The pickup bound is conservative, especially when divergence is large and historical pickup is small at the chosen horizon. Users who require parametric forecasts of the full lead time vector can adopt B\textendash DARMA or related compositional time series models while still benefiting from divergence monitoring and the risk index \citep{KatzBruschWeiss2024}. The decision templates are intentionally simple and are intended as safe defaults that can be tuned over time.

\section{Conclusion}
By turning two ideas from the article into a minimal R artifact that analysts can run on their own files, it becomes easier to detect distributional change, to translate that change into a clear risk index, and to connect the index to concrete actions. The same artifact defines how to measure effects in operational settings. This closes the loop between research and decision and sets the stage for comparative evaluation across markets and segments.

\section*{Declarations}
\textbf{Funding} None declared.

\noindent\textbf{Competing interests} The authors declare no conflicts of interest and that all work and opinions are their own and that the work is not sponsored or endorsed by Airbnb.

\noindent\textbf{Data availability} All results are generated from simulations. A synthetic dataset and an R package archive are provided as supplementary material.

\noindent\textbf{Code availability} The R package \texttt{leadtimefluxR} and example scripts are provided as supplementary files and are available at \url{https://github.com/harrisonekatz/leadtimefluxR}.

\noindent\textbf{Ethics approval, consent, and permissions} Not applicable.

\end{document}